\documentstyle[aaspptwo]{article}
\tighten          
\slugcomment{submitted to ApJ {\em Letters\/}}
\newcommand{\beq}{\begin{equation}}
\newcommand{\eeq}{\end{equation}}
\newcommand{\beqa}{\begin{eqnarray}}
\newcommand{\eeqa}{\end{eqnarray}}

\newcommand{\teffa}{T_{\rm eff} }

\newcommand{\ea}{{\it et al. }}

\newcommand{\Li}{\hbox{$^7$Li}}
\newcommand{\be}{\hbox{$^9$Be~}}

\newcommand{\li}{\hbox{$^7$Li~}}
\newcommand{\lisix}{\hbox{$^6$Li~}}

\newcommand{\Lie}{{\cal L}_{\lower4pt\hbox{$\!\!\!\!\scriptstyle\xi$}}}
\newcommand{\feh}{[{\rm Fe}/{\rm H}]}
\newcommand{\nli}{\log {\rm N} ({\rm Li})}

\begin{document}

\title{The Primordial Abundance of $^6$Li and $^9$Be}

\author{Brian Chaboyer}

\affil{Candian Institute for Theoretical Astrophysics,
University of Toronto, \\
60 St. George Street,  Toronto, Ontario, Canada  M5S 1A7\\
Electronic Mail: chaboyer@cita.utoronto.ca}

\begin{abstract}
Light element ($^6$Li, \li and $^9$Be) depletion isochrones for halo
stars have been calculated with standard stellar evolution models.
These models include the latest available opacities and are computed
through the sub-giant branch.  If \lisix is not produced in
appreciable amounts by stellar flares, then the detection of
\lisix in HD 84937 by Smith, Lambert \& Nissen (1993) is compatible
with standard stellar evolution and standard big bang nucleosynthesis
only if HD 84937 is a sub-giant.  The present
parallax is inconsistent with HD 84937 being a sub-giant star at the
$2.5\, \sigma$ level.  The most metal poor star with a measured \be
abundance is HD 140283, which is a relatively cool sub-giant.
Standard stellar evolution predicts that \be will have been depleted
in this star by $\sim 0.3$ dex (for $\teffa = 5640$ K).
Revising the abundance upward changes
the oxygen-beryllium relation, suggesting  incompatibility with standard
comic ray production models, and hence,  standard big bang
nucleosynthesis.  However, an increase in the derived temperature
of HD 140283 to 5740 K would result in little depletion of \be and
agreement with standard big bang nucleosynthesis.

\end{abstract}


\vspace*{20pt}
\begin{center}
{\large submitted to {\it Astrophysical Journal Letters\/}}
\end{center}

\twocolumn

\section{Introduction}
The primordial abundance of the light elements provide a crucial
test of standard big bang
nucleosynthesis (SBBN) theory.  In SBBN,  the abundances of the light
elements are functions of the baryon to photon ratio
in the early universe.  SBBN does not produce appreciable amounts of
$^6$Li, or any element heavier than \Li.
Consistency with the inferred primordial abundance of $^4$He
 and (D + $^3$He) requires that the baryon to photon ratio
($\eta$) lies in the narrow range
$2.86\times 10^{-10} \le \eta \le 3.77\times 10^{-10}$ and
implies that the primordial \li  abundance should be
$1.9 \le \nli \le 2.4$ (Smith, Kawano \& Malaney 1993, where
$\nli = \log ({\rm Li/H}) + 12$).  The primordial
$^4$He abundance is restricted to the range $0.237 \le Y_p \le 0.247$.
Various alternative to SBBN have been proposed, some of which
predict significantly different values for the primordial $^6$Li, \li
and \be
abundances (cf. the review by Malaney \& Mathews 1993).
An accurate determination of the primordial light element  abundances
has important implications for particle physics and cosmology.
However, $^6$Li, \li and \be are fragile elements which can be
destroyed within stars.  In addition, these elements may be
produced by cosmic rays in the early galaxy.
It has also been suggested that stellar flares may produce observable
amounts of \lisix in stars which contain very thin convective
envelopes (Smith, Lambert \& Nissen 1993, hereafter SLN93; Deliyannis
\& Malaney 1994)
In order to relate the presently observed abundances to their
primordial value, it is important to study possible
depletion/production mechanisms which occur after the big bang.

The observed abundances of the light elements in halo
stars\footnote{The term halo star is used to denote extremely
metal poor stars ($\feh < -1.0$), which are likely to be
the oldest stars in our galaxy.} provide our best starting point for
determining the primordial abundances, as these low metallicity stars
are the least affected by post big bang production mechanisms.
It was originally shown by Spite \& Spite (1982) that halo
stars  with
effective temperatures greater than 5700 K have a nearly uniform
Li abundance\footnote{Observers are usually not able to measure the
isotopic abundance of Li, so they measure the total Li content in
a star, which we denote by Li.  However the theoretical calculations
predict considerably different production/depletion for the two stable
Li isotopes, \lisix and \Li, and so our discussion of theoretical work
will retain the distinction between \lisix and \Li.}.
The consistency of the Li
abundance in hot halo stars was confirmed by other observations (eg.~Spite
\& Spite 1986; Hobbs \& Thorburn 1991).
The observed Li abundance ($\nli \simeq 2.1$)
is consistent with the predictions of SBBN ($1.9 \le \nli \le 2.4$,
Smith \ea 1993).  In standard stellar
evolution theory the only way in which the observed surface value of a light
element  would be depleted
is if the convection zone is deep enough to dredge up material which
has destroyed the light element.  Detailed modeling by
Deliyannis, Demarque \& Kawaler (1990) demonstrated that the
convection zones in old, metal poor stars are too shallow to dredge up
substantial amounts of \li depleted material, hence the
observed \li abundance does indeed reflect the primordial value
(provided post big bang enrichment is neglible).
However, if mixing occurs within the radiative region of a star, then
substantial \li depletion may occur.  Models which include diffusion
find primodial $^7$Li abundances 0.2 dex higher than standard models
(Deliyannis \ea 1990; Proffitt \&
Michaud 1991; Chaboyer \& Demarque 1994).  Instabilities induced by
rotation may also mix material in stellar radiative regions.  Models
which include rotation induced mixing found \li depletion of $\sim
0.9$ dex, and imply a primordial \li abundance incompatible with SBBN
(Pinsonneault, Deliyannis, \& Demarque 1992; Chaboyer \& Demarque 1994).

The predictions of SBBN
may be put to a more stringent test if we consider other light
elements.  \be has been observed in a substantial number of stars
(eg. Gilmore \ea 1992, hereafter GGEN92; Boesgaard \& King 1993,
hereafter BK93), while B has been observed in a few stars
(Duncan, Lambert \& Lemke 1992; Edvardsson \ea 1994).
These elements require substantially
higher temperatures to be destroyed, and so are less likely to be
depleted in stellar interiors.  SLN93 have recently claimed to have
detected \lisix in a single halo star, HD 84937.  This detection
awaits confirmation, but it is assumed here that the detection is
real.  This {\it Letter\/}
will examine the light element depletions predicted
by standard stellar evolution and compare them to the observations.

\section{Li Depletion Isochrones}
Light element depletion in halo stars has been studied by
Deliyannis \ea (1990); Proffitt \& Michaud (1991); Deliyannis \&
Demarque (1991) and Pinsonneault, Deliyannis \&
Demarque (1992).
The primary difference between the standard models presented here, and
those in earlier studies
is the opacity used in the models. The present models use the opacities of
Iglesias \& Rogers (1991) for high temperatures and Kurucz (1991)
opacities for temperatures below $10^4$ K.  Previous work used
opacities from Cox \& Stewart (1970), or
Huebner \ea (1977).  The new opacities are substantially enhanced over the
old values in certain temperature, density regimes which leads to
deeper convection zones in most of the models.

An extensive grid of standard stellar models, with
$\rm M = 0.55~M_\odot - 0.82 ~M_\odot$ (in $0.01\rm ~M_\odot$
intervals) and $\rm Z = 10^{-3},~10^{-4}$ and $10^{-5}$ has been
constructed in order to study the light element depletion in halo
stars.  These models are evolved from the fully convective pre-main
sequence to an age of 18 Gyr, or through the sub-giant phase of
evolution, whichever occurred first.  Full details of the model
construction may be found in Chaboyer \& Demarque (1994).
These models were used to construct
light element destruction isochrones with  ages of
14 and 17 Gyr, chosen to span the typically age range observed in the halo
(eg. Chaboyer, Sarajedini, \& Demarque 1992).
These isochrones are available in electronic
form from the author.  Chaboyer \& Demarque (1994) compared
the \li isochrones to the observations and found good agreement
with the observations of Thorburn (1994) implying a primordial
\li abundance of $\nli = 2.24\pm 0.03$.

The Li destruction isochrones are presented in Figure \ref{fig2}
for $t=17$ Gyr, $Z=10^{-4}$.  Above $\sim 5900$ K, the
\lisix depletion is constant on
the {\it sub-giant\/ branch}, with a depletion factor (defined as
$\rm D \equiv ~ ^6Li/^6Li_{protostellar}$) of approximately 0.3.
\begin{figure}[t]
\vspace*{6cm}
\caption{Standard Li destruction isochrones at age of 17 Gyr with
$Z=10^{-4}$ on the main sequence and during sub-giant evolution.
The vertical axis is defined as $\log \rm D \equiv \log
(Li/Li_{protostellar})$.
The two lower lines are for $^6$Li, while the two upper
lines are for $^7$Li.}
\label{fig2}
\end{figure}
However, no such depletion plateau occurs for \lisix on the main
sequence.  The predicted abundance of \lisix is a strong function of
effective temperature on the main sequence.  The
sub-giant \li isochrone is a nearly constant offset from the sub-giant
\lisix sub-giant isochrone.  Thus, the standard models predict
observations of the $\rm ^6Li/^7Li$ ratio
should be constant over a wide range of effective temperatures in halo
sub-giant stars.

The light element depletion isochrones are sensitive to the mixing
length used in the models (Deliyannis \ea 1990).  The models presented
here use a solar calibrated mixing length ($\alpha = 1.728$).
These models were
used to construct isochrones, and compared to the colour-magnitude diagram of
M15 (Durrell \& Harris 1993), a metal-poor globular cluster.
Conversion from the theoretical temperature-luminosity plane to the
observed colour-magnitude plane was performed using both the Green
(1988) colour calibration and the Kurucz (1991) colour calibration.
Both sets of
isochrones are a good fit\footnote{In performing the isochrones fits,
the reddening and distance modulus were allowed to vary within twice
the quoted errors of Durrell \& Harris (1993).  A fit was deemed
satisfactory if the isochrones were able to simultanesouly match the
location of the lower main sequence and sub-giant/giant branches.}
to the observations.  Hence, the effective
temperatures derived from the models are in good agreement with
globular cluster observations.   Stellar models and isochrones
were also constructed with mixing lengths of $\alpha = 1.5$ and 2.0.
It was found that most of these isochrones did not match the colour magnitude
of M15, indicating that the solar calibrated mixing length is the
appropriate one to use for halo stars.  However, the $\alpha = 2.0$ isochrone
with the Kurucz colour calibration did provide a satisfactory fit.
This suggests that the
effective temperatures predicted by the stellar models are accurate to
within $\sim 80$ K on the main sequence and sub-giant branches.

\section{The Case of HD 84937}
To date, \lisix has been claimed to be detected in a single halo star
-- HD 84937 by SLN93.  They determined $\rm ^6Li/^7Li = 0.05\pm 0.02$
and $\nli = 2.12$. In this {\it Letter\/} the actual $\rm ^6Li/^7Li$
ratio in HD 84937 is assumed to lie in the range $0.03 - 0.07$.  SLN93
determined the  metallicity to be $\feh = -2.4$ and
$\teffa = 6090 \pm 100$ K.  This is $\sim 100$ K lower than
other estimates for this star.
For example, Thorburn (1994) determined an effective
temperature of $6232 \pm 100$ K for this star.  King (1993) advocates
a higher temperature scale for halo stars, and finds $\teffa = 6312$.
We will assume that the effective temperature of HD
84937 lies in the region $6000 - 6300$ K.  The evolutionary status of
HD 84937 is also somewhat uncertain.  Str\"{o}mgren photometry locates
HD 84937 at the bluest point of the turnoff (SLN93).  However, the
effective temperature of this star is too low for it to be a massive main
sequence star.  Stellar models with the metallicity of HD 84937
typically have turn-off temperatures ranging from 6700 K (age 14
Gyr) to 6540 K (age 18 Gyr).  SLN93 suggest that HD 84937 is a
sub-giant, however it is possible that HD 84937
is a lower mass main-sequence star.
We will consider both possibilities in the discussion which follows.

Our $Z = 10^{-4}$ isochrone corresponds to $\feh = -2.55$ and
$[\alpha/{\rm Fe}] = +0.40$ (assuming that the effects of
$\alpha$-element enhancement may be accounted for by modifying the
overall $Z$ of the star, as outlined by Salaris, Chieffi \& Straniero 1993)
and so is the most appropriate
isochrone to compare to the observation.  If HD 84937
is a main sequence star, then the models predict that the protostellar
\lisix has been depleted by a factor of 13 ($\teffa = 6300$ K) to 250
($\teffa = 6000$ K).
The \li is not depleted in the models over this
effective temperature range.  Hence, if HD 84937 is a main sequence
star, then the {\it protostellar\/} $\rm ^6Li/^7Li$ ratio is in the range
$0.40 - 17$.  This requires a large primordial abundance of \lisix or
substantial cosmic ray production of $^6$Li.  The
$\rm ^6Li/^7Li$ ratio produced by cosmic rays is approximately one,
thus production of large amounts of \lisix by cosmic rays
would produce substantial amounts of \Li.  This would lead to a
correlation between metallicity and Li abundance in the
halo stars, which is not observed.  Thorburn 1994 finds a correlation
which is a factor of 2 too small to account for a protostellar
$\rm ^6Li/^7Li$ ratio of 0.40. Thus, consistency with SBBN and
standard stellar evolution models implies that HD 84937 is not a main
sequence star.  In analysing the \lisix detection in HD 84937,
Steigman \ea (1993) arrived at the opposite conclusion, namely that
the detection was consistent with standard models, even if HD 84937 is
a main sequence star.  The difference between Steigman \ea (1993) and
this work is that Steigman \ea (1993) used the standard stellar
models of Deliyannis \ea (1990) which deplete less \lisix then the
models presented here.  In addition, Steigman \ea (1993) allowed for
a much larger error in the observed \lisix abundance.

If HD 84937 is a sub-giant, then the situation is considerably
different.  For effective temperatures between 6000 and 6300 K,
the sub-giant models predict that \lisix is depleted by a
factor of 2.8, while virtually no depletion of \li occurs.  This
relatively mild amount of \lisix depletion implies that the
protostellar $\rm ^6Li/^7Li$ ratio is in the range $0.084 - 0.20$.
This range is consistent with SBBN (which produces no
$^6$Li) and production of \lisix and \li (in roughly equally
proportions) by cosmic rays. Thus, the observations are
consistent with SBBN,  provided that HD 84937 is a sub-giant.
The parallax of HD 84937 is  $0.0277''\pm 0.0065''$
(van Altena, Lee \& Hoffleit 1994).
If HD 84937 is a sub-giant, then the models
predict a minimum absolute $V$ magnitude of $M_V = 3.56$
(corresponding to a 17 Gyr sub-giant with $\teffa = 6300$ K).
The apparent $V$ magnitude of HD 84937 is observed to be $V = 8.32$
(Stetson \& Harris 1988). This
implies a parallax of $0.0112$, which is $2.5\, \sigma$ smaller than
observed.  If HD 84937 is a main sequence star, then the models
predict a parallax ranging from $0.0205''$
to $0.0273''$, which agrees with the observed
value.  We caution however, that  measurement of a parallax is a
difficult observations, and so the true error
in the parallax may be higher than the quoted value.
An improved determination of the parallax (possible with HIPPARCOS) would be
able to answer the question of the evolutionary status of HD 84937,
and so would serve as a test of the `standard model' (standard
stellar evolution, no production of \lisix by stellar flares,
standard production of light elements via cosmic rays, and SBBN).

\section {$^9$Be Destruction Isochrones}
The \be destruction isochrones in standard models are shown in Figure
\ref{figbeLA} for $Z=10^{-4}$ (the $Z = 10^{-5}$ and $Z = 10^{-3}$
isochrones are virtually identical).  No \be destruction occurs on
the main sequence, but \be destruction occurs on the sub-giant branch
when $\teffa \la 5750$ K.
\begin{figure}[t]
\vspace*{6cm}
\caption{Standard $^9$Be destruction isochrones at age of 17 Gyr with
$Z=10^{-4}$ on the main sequence and during subgiant evolution.
$^9$Be destruction occurs for sub-giants with $\teffa \la 5700$ K.}
\label{figbeLA}
\end{figure}
This can have important implications for the
interpretation of \be observations. For example, the most metal poor
star observed is HD 140283 with $\feh = -2.77$ (GGEN92).
GGEN92 determined a \be abundance of
$\log \ (N_{\rm Be}/N_{\rm H}) = -12.97\pm 0.25$
and  quote $\teffa = 5540$ K,  $\log g = 3.5$.
BK93 find $\log \ (N_{\rm Be}/N_{\rm H}) = -12.78\pm 0.14$, $\teffa = 5660$ K
and $\log g = 3.6$.  Ryan \ea (1992) obtained
$\log  (N_{\rm Be}/N_{\rm H}) = -13.25\pm 0.4$ with $\teffa  = 5700$ K, and
$\log g = 3.2$.    These effective temperatures and
gravities imply  that HD 140283 is a sub-giant, which has
depleted $^9$Be according to standard stellar evolution models.  The
amount of \be depletion in stellar models is extremely sensitively
to the assumed effective temperature.
The depletion factor is $0.46$ dex at $\teffa = 5585$ K,
$0.2$ dex at $\teffa = 5660$ K and
$0.1$ dex at 5700 K.  Thus, the protostellar \be abundance in HD
140283 inferred from the observations is
$\log  (N_{\rm Be}/N_{\rm H}) =-12.51$, $-12.58$ and $-13.15$
(GGEN92, BK93, and  Ryan \ea 1992 respectively).

Revising the \be abundance of HD 140283 upward changes the
oxygen-beryllium relation, as HD 140283 is the most metal poor star
observed.  The abundance of \be in other metal poor stars
is not affected, as these stars are hotter than HD 140283.
If we use the GGEN92 or BK93 abundance, then
HD 140283 would have a similar or higher \be abundance than HD 213657,
HD 106617 and HD 116064 which have oxygen abundances 0.4 dex higher
than HD 140283.  This suggests that there may exist a plateau in
the \be abundance at low  metallicity, implying a
primordial origin to the $^9$Be.  This is incompatible with SBBN.
A least squares analysis of the O--\be data of GGEN92 for stars with
$\rm [O/H] \le -1.0$
yields a correlation coefficient of 0.57, and a slope of 0.5.
This slope is inconsistent with
the simple cosmic ray production models of GGEN92 which predict a
slope of 1 -- 2.

The low \be abundance found by Ryan \ea (1992) is
compatible with simple cosmic ray production models (e.g.~ Steigman
\ea 1993).
If the actual temperature of HD 140283 is
$\teffa \ga 5740$ K, then no \be depletion is predicted in the
standard stellar evolution models, and agreement is found with SBBN.
King (1993) advocates a higher effective temperature scale for halo
stars, and determines $\teffa = 5812$ K for HD 140283.  At such
temperatures, no \be depletion is predicted by standard stellar
evolution models, and evidence for a plateau in the \be abundance is
weakened.  As noted earlier, there is a possible error of $\sim 80$ K
in the models.  Thus, the exact amount of \be depletion in cool
sub-giants is somewhat uncertain.  Observing stars which are hotter
than $\teffa = 5800$ K removes this uncertainty.

Edvardsson \ea (1994) have measured the B abundance in HD 140283 to be
$\log \ (N_{\rm B}/N_{\rm H}) = -11.66\pm 0.2$ when a large (+0.54 dex)
non-LTE correction is applied.  From this, they infer an abundance
ratio of $N_B/N_{Be}$ of 17.  This is in good agreement with the ratio
10 -- 20 predicted by cosmic ray production (Duncan \ea 1992; Steigman
\& Walker 1992).   If the \be abundance is corrected for the
$\sim 0.3$ dex depletion implied by the stellar models, (no B depletion is
predicted for this star) then the abundance ratio is
$N_B/N_{Be} = 9$, with a range of 4 -- 20 (assuming a 0.3 dex error
in the ratio due to uncertainties in the observations, and 0.2 dex due
to uncertainties in the depletion).
More observations of \be in extremely metal
poor stars and a better understanding of the effective temperature
scale in halo stars are needed to definitively settle the question of the
primordial $^9$Be abundance.

\section{Summary}
Surface depletion isochrones for $^6$Li, \li and \be have been
constructed for $Z=10^{-3},~10^{-4}$ and $10^{-5}$, $t = 14$ and 17
Gyr from
standard stellar evolution models using the opacities of
Iglesias \& Rogers (1991) and Kurucz (1991).
The isochrones include sub-giant branch evolution and are available in
electronic form from the author.
The observed detection of \lisix in HD 84937 by SLN93 is compatible
with the ``standard'' model for the production of light elements
(SBBN, standard cosmic ray production, no stellar flare production,
standard stellar evolution) only if HD 84937 is a sub-giant.  The
observed parallax rules out this possibility at the $2.5\, \sigma$
level.  A more accurate parallax measurement would serve as a test of
the standard model for light element production.  The models predict
that the abundance of \lisix should be a strong function of effective
temperature on the main sequence. Observations of \lisix in a
number of main sequence halo stars would be a good test of these models.

Only 8 halo stars with $\rm [O/H] \le -1$ have observed \be
abundances (GGEN92; BK93). The
most metal-poor of these, HD 140283 is a relatively
cool sub-giant.  Standard stellar evolution models predict that
\be is depleted in this star, by $\sim 0.3$ dex
(for $\teffa = 5640$ K), implying a protostellar
\be abundance of $\log  (N_{\rm Be}/N_{\rm H}) =-12.54$ (GGEN92;
BK93).  The exact amount of \be depletion is  very sensitive to the
effective temperature.  Ryan \ea (1992) determined a lower
\be abundance (and higher $\teffa$).  If we use the results of GGEN92 and BK93
then revising the abundance upward changes the oxygen-beryllium relation,
suggesting a plateau in the beryllium abundance.
This is incompatible with standard cosmic ray production
and SBBN.  An increase in the observed effective temperature of HD
140283 by $\sim 100$ K (as suggested by King 1993) or the use of the
Ryan \ea (1992) abundance resolves  this discrepancy.  It is clear
that an accurate effective temperature and $\log g$
(critical in determining O and Be abundances) scales for halo stars
are crucial for our understanding of the origin of the light
elements.

\acknowledgments
I would like to thank R. Malaney for numerous discussions and the
anonymous referee, whose comments improved the presentation of this paper.
This research has made use of the Simbad database, operated at CDS,
Strasbourg, France.

\appendix


\end{document}